\documentclass[rnote]{aa} 
\usepackage{time}
\usepackage{color}
\usepackage{natbib}
\usepackage[percent]{overpic}
\usepackage{amstext,amsmath,mathtools}
\usepackage{graphicx}
\usepackage{textcomp}
\usepackage{txfonts}
\usepackage{url}

\begin{document}
\titlerunning{Response to ``Stray-light correction in 2D spectroscopy''}
\title{Response to ``Stray-light correction in 2D spectroscopy'' by R.~Schlichenmaier and M.~Franz}

\author{G.B. Scharmer\inst{1,2,3}}

\institute{Institute for Solar Physics, Stockholm University,
AlbaNova University Center, SE 106\,91 Stockholm, Sweden \and
Stockholm Observatory, Dept. of Astronomy, Stockholm University,
AlbaNova University Center, SE 106\,91 Stockholm, Sweden \and
Royal Swedish Academy of Sciences, Sweden}
\date{Draft: \now\ \today}
\frenchspacing

\abstract{We discuss a recent paper by Schlichenmaier \& Franz (SF; 2013, A\&A, 555, A84), in which the claim is made that the penumbral dark downflows detected for the first time with the Swedish 1-m Solar Telescope (SST) by Scharmer et al. and Joshi et al. could be produced by overcompensation for straylight. We show that the analysis of SF is fundamentally flawed, because it ignores the constraints on the strength of such straylight from 3D convection simulations and on the spatial extent of the straylight point spread function from the measured minimum intensity in the sunspot umbra. Furthermore, we show that the claim made by SF, that the spatial straylight of Hinode is less than 10\%, is false. We conclude that the analysis of SF is of no relevance in relation to the straylight compensation method applied to the SST data. 
}


\maketitle

\section{Introduction}
After more than 100 years of spectroscopic observations, we finally have the tools to fully explain the origin of sunspot fine structure such as bright dots in the dark umbrae and the penumbral filamentary structure and strong radial outflows, discovered by \citet{1909MNRAS..69..454E}. This progress can only be partly attributed to improvements in spatial resolution of solar telescopes, the development of spectropolarimeters capable of providing 3D views of properties of solar fine structure, and the inversion techniques developed for interpreting such data. Of equal importance is that our ability to model these structures in a realistic way through numerical simulations has developed dramatically during just the last 6 years. 

Simulations now uniquely attribute the origin of the above sunspot fine structure and the Evershed flow to convective processes operating below the visible surface in strong magnetic fields that are either nearly vertical (in the umbra) or more horizontal (in the penumbra). However, there is not yet full consensus as regards whether or not the interpretation of sunspot penumbrae implied by simulations is actually supported by observations. In particular, \citet{2009A&A...508.1453F,2013A&A...550A..97F}, find (patchy) downflows mainly in the mid and outer penumbra, supporting an interpretation in terms of arching magnetic flux tubes. Recently, observations have been presented that for the first time report on both convective upflows and (in particular) downflows in the entire penumbra \citep{2011Sci...333..316S, 2011ApJ...734L..18J, 2013arXiv1307.3668T}, consistent with theoretical predictions \citep{2006A&A...447..343S, 2006A&A...460..605S} and simulations \citep{2007ApJ...669.1390H, 2009Sci...325..171R, 2009ApJ...691..640R, 2011ApJ...729....5R}. These simulations also show azimuthal convective flows in addition to the vertical up/down flows and the outward Evershed flow. 

In a recent paper \citet{2013A&A...555A..84S}, hereinafter SF, cast doubt on the straylight compensation employed by \citet{2011Sci...333..316S} and \citet{2011ApJ...734L..18J}, and conclude that the red shifts detected ``could'' be artifacts produced by overcompensation for straylight. Thus they would not be genuine downflows. However, we show below that this conclusion is based on a simplistic analysis that ignores the constraints on the straylight compensation set by 3D convection simulations and the measured umbra intensity, employed by  \citet{2011Sci...333..316S}. Furthermore, the calculations made by SF, intended to demonstrate false downflow detections from overcompensation of straylight, are coupled to the assumption that the true straylight of Hinode spectropolarimetric data ``clearly'' is below 10\%. This statement is false, as we will demonstrate below. In Sect.~2, we discuss the straylight compensation of \citet{2011Sci...333..316S}.
In Sect.~3, we discuss the point spread function of Hinode and demonstrate that it leads to spatial straylight of at least 35\%. Finally, in Sect.~4, we point out additional misleading statements and analysis made by SF, and in Sect.~5, we summarize our results and findings.

\section{Actual constraints of the straylight compensation method employed}
The most serious problem with the analysis of SF is that it creates the illusion of addressing the straylight problem in the same way as \citet{2011Sci...333..316S}, while in fact the analysis of SF deals with a much less constrained problem. By failing to address the relevant problem, their analysis and argumentation is fundamentally flawed and thus useless in supporting any of their conclusions. This is quite disturbing, since the method used for straylight compensation by Scharmer et al. as well as the associated simulations and analysis leading up to this particular choice are well documented \citep[][their Supporting Online Material, SOM\footnote{{\small \url{http://www.sciencemag.org/content/333/6040/316/suppl/DC1}}}]{2011Sci...333..316S}. 

The main assumption used by \citet{2011Sci...333..316S} to constrain the straylight point spread function (PSF) is that numerical simulations predict the correct root mean square (rms) continuum intensity contrast. 
This assumption receives strong support in measurements of granulation contrast made with data from Hinode and compared to that of synthetic granulation images calculated from 3D simulations \citep{2008A&A...487..399W, 2008A&A...484L..17D, 2009A&A...503..225W, 2009A&A...501L..19M}. In particular, \citet{2009A&A...503..225W} show that deconvolved Hinode images have a contrast that is consistent with the CO$^5$BOLD, MURaM and Stagger codes. This close congruence of the predicted granulation contrast from 3 independently developed simulation codes is confirmed in independent work \citep[][their Table 4]{2012A&A...539A.121B}. The work of \citet{2013A&A...554A.118P} further establishes confidence in the 3D convection simulations by demonstrating superior agreement of all investigated diagnostics based on a 3D hydrodynamic model obtained with the Stagger code, when compared to any of the tested 1D models. There is thus overwhelming support from theoretical simulations and observations for the main assumption of the method used by \citet{2011Sci...333..316S} to constrain the straylight PSF. 

Our synthetic granulation images degraded by the diffraction limited PSF of the Swedish 1-m Solar Telescope \citep[SST;][]{2003SPIE.4853..341S} has a contrast of 16.9\% at 538~nm, whereas the observed contrast at that wavelength is only 8.9\% (SOM, Table S1). Similarly, the calculated synthetic granulation rms velocity is 1.78~km~s$^{-1}$, whereas the corresponding measured velocity is only 0.84~km~s$^{-1}$, consistent with the level of contrast degradation of the observed granulation images. This comparison immediately suggests that there must be on the order of 50\% spatial straylight in the observed data. This is because granules have typical sizes of 1\farcs5, and only the wings of the PSF of a 1-meter telescope can achieve such contrast reduction.

To set an upper limit on the spatial extent of the straylight PSF, we take advantage of the relatively dark sunspot within the same field-of-view (FOV) as the observed granulation pattern used to constrain the strength of the straylight PSF. The minimum umbra intensity is only 15.6\% of that of the surrounding granulation (SOM, Table S2), excluding the possibility that the dominant part of the straylight PSF is very wide. In particular, this in practice excludes a PSF with purely Lorentzian wings. Performing instead the analysis with Gaussian straylight profiles allows good fits to the synthetic data, but if the full width at half maximum (FWHM) of the straylight PSF is significantly larger than 2\farcs4, the minimum umbra intensity of the straylight compensated data is negative. 

Our imaging model can be expressed with the following equation, relating the observed $I_o$ and ``true'' $I_t$ intensities: 
\begin{equation} 
I_o=[(1-\alpha)~\delta + \alpha~G(W)] * T * I_t, 
\end{equation}
where $T$ represents the PSF of the telescope, calculated from low-order aberrations originating in the Earth's atmosphere, the telescope and the instrumentation used, and $\delta$ represents the (spatial) delta function. The parameter $\alpha$ is the straylight fraction, `` * '' denotes convolution, and $G$ is a Gaussian straylight point-spread function (PSF), having a full width at half maximum of $W$. Note, that the image reconstruction preceding the straylight compensation corresponds to the compensation for the telescope PSF $T$, such that the stray-light compensation only involves the PSF contained within the square brackets of Eq. (1).

Our final choice was a Gaussian with a FWHM of 1\farcs2 and a straylight fraction of 58\%. This gave an rms continuum intensity of 17.0\% and velocity of 1.8~km~s$^{-1}$ for the straylight compensated data, matching the values of the simulations almost perfectly. The minimum umbra intensity obtained with this choice is 11.1\%. We note (SOM, Table S2) that a FWHM of 2\farcs4 and a straylight fraction of 50\% would fit the simulated data nearly as well, but would reduce the minimum umbra intensity by a factor 3 from the observed value to 5.6\%. These values obtained for the minimum umbra intensity with straylight compensation are reasonable. For example, \cite{2007A&A...465..291M} obtained minimum umbra intensities in the range 5--20\% with an average of 12\% for sunspots with umbra diameters similar to that of our observed sunspot (about 10\arcsec), using straylight compensated MDI data recorded at 677~nm.

There are clearly unavoidable uncertainties associated with the proper determination of the straylight PSF from these SST data. In particular, the minimum umbra intensity does not allow a precise determination of the FWHM of the straylight PSF. Nevertheless, our analysis shows that the main conclusion of the paper, which is the {\em detection} of dark penumbral downflows, is a very robust result. With 58\% straylight and a FWHM of 1\farcs2, the average redshifts of dark downflows differ by only 100~m~s$^{-1}$ from those obtained assuming 50\% straylight and a FWHM of 2\farcs4. In fact, whether we use a Gaussian or even Lorentzian straylight PSF, the dark penumbral structures show an average downflow velocity of at least several hundred m~s$^{-1}$, as long as the straylight compensation is such that it leads to rms intensity and velocity fluctuations that are similar to those predicted by the simulations (SOM, Sect.~1.2), degraded to the SST diffraction limited PSF.

A relevant question is certainly whether there exists a different choice of straylight PSF that provides consistency with the simulated rms intensity and velocity and leads to a minimum intensity in the umbra that is larger than zero, and yet invalidates our detection of dark downflows. We firmly believe that this is not the case. This is however not the question addressed by SF. 

\section{Hinode straylight}
SF based their analysis on straylight compensated Hinode data, using a PSF identical to that within the square brackets of Eq. (1) but with a FWHM corresponding to 4\farcs7, which is nearly 4  times larger than used by \citet{2011Sci...333..316S}. With $\alpha$=0.6, this leads to rms velocity fluctuations that are about 2.5 times larger than for uncorrected data for quiet Sun granulation and 2--2.5 times larger within the penumbra (SF, Tables 1 and 2).

Our second serious objection to the paper of SF, however, is that the authors claim that the true straylight ($\alpha$-value) ``clearly'' is less than 10\% for Hinode. This misleads the reader to conclude that their straylight compensation results in false detections of downflows and therefore provides a link in their chain of evidence supporting their conclusions.

We show below that the straylight of Hinode should be considered to be at least 35\% and could even be as high as 40\%. To do this, we follow the approach of modeling the Hinode PSF of  \citet{2008A&A...484L..17D}, referred to as the source for the claim that the straylight is less than 10\% by SF. This model of the Hinode PSF is also employed in the inversions discussed by \citet{2013arXiv1307.3668T}. 

Danilovic et al. start off from a simulated granulation image. Degrading this with the PSF of a diffraction limited un-obscured 50-cm telescope reduces the rms intensity contrast at 630~nm from 14.4\% to 10.9\%. Taking into account the central obscuration, the spider, the CCD pixel size and a focus error of 1.5~mm degrades the rms contrast to 7.5\%. Danilovic et al. also mention that (additional) straylight of only 4.7\% would reduce the rms contrast to 7.0\%, which is close to the observed contrast and probably the origin of the statement by SF. 

To explain this strongly reduced granulation contrast in Hinode data, we calculate the modulation transfer function (MTF) \citep[calculated also by][and shown in their Fig.~1]{2008A&A...484L..17D} and the corresponding PSF. Table 1 summarizes the models used to represent different aspects of the Hinode Optical Telescope Assembly \citep[OTA;][]{2008SoPh..249..197S}. These follow closely \citet{2008A&A...484L..17D}, except that we have not included the MTF corresponding to the CCD pixel size and we have not been able to translate the 1.5~mm defocus into a precise rms focus error. Instead we have assumed this to correspond to a defocus of 40~nm rms (counted within the unobscured part of the pupil), which seems to give reasonable agreement with the MTF calculated by Danilovic et al. The cases studied here correspond to: a diffraction-limited unobscured 50-cm aperture (Case I), the same aperture with an added 34.4\% central obscuration (Case II), three 40~mm wide spiders (Case III), and with 40~nm rms focus error added (Case IV). The pupil geometry was adopted from \citet{2008SoPh..249..197S}. 

\begin{table}
  \centering
  \caption{Hinode point spread function model and calculated properties.}
  \begin{tabular}{ l l r r r}
    \hline\hline\noalign{\smallskip}
    Case & Degradation & Strehl & E$_{core}$ & $\Delta E_{wing}$\\
    \hline\noalign{\smallskip}
     I & 50 cm aperture& 1.00 & 0.84 & 0.00 \\
     II &+ Central obscuration& 0.88 & 0.64 & 0.20  \\
    III & + Spider& 0.78 & 0.55 & 0.29  \\
    IV & + 40 nm rms defocus& 0.66 & 0.49 & 0.35 \\
    \hline
  \end{tabular}
  \tablefoot{Summary of models and calculated properties of the corresponding point spread functions (PSF's) for the 4 cases studied. The Strehl is the peak intensity of the PSF, normalized to that of Case I, E$_{core}$ the fraction of encircled energy within the diffraction core (inside the first diffraction minimum), and $\Delta E_{wing}$ the fraction of encircled energy added to the diffraction rings (i.e., outside the diffraction core) for Case II-IV, relative to that of Case I.}
\label{tab:table_cd}
\end{table}

\begin{figure*}[tbp]
\center
\begin{overpic}[bb=55 70 570 710,width=0.3\textwidth,angle=-90]{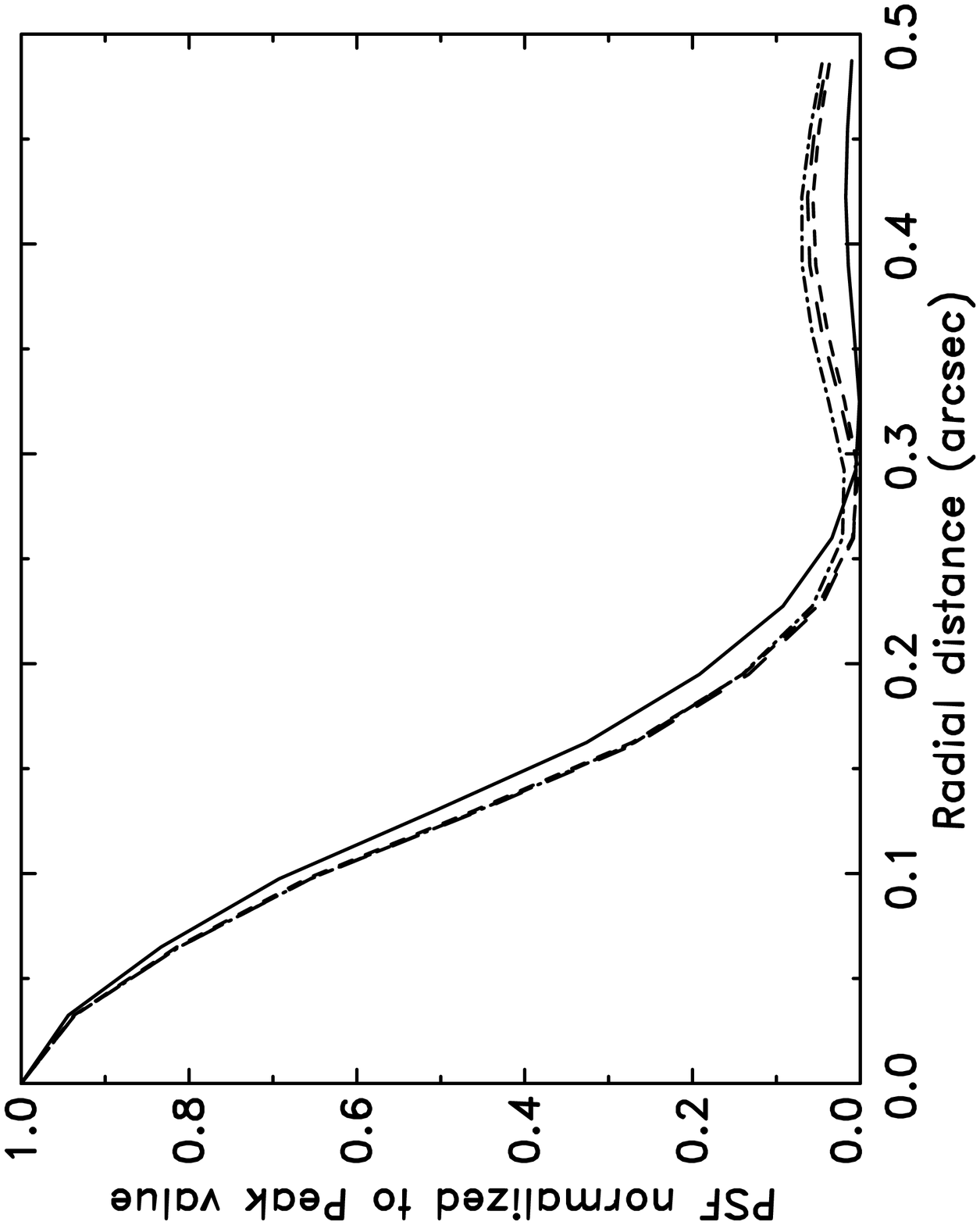}
\put(87,63){\textcolor{black}{\textbf{a)}}}
\end{overpic}
\begin{overpic}[bb=55 70 570 710,width=0.3\textwidth,angle=-90]{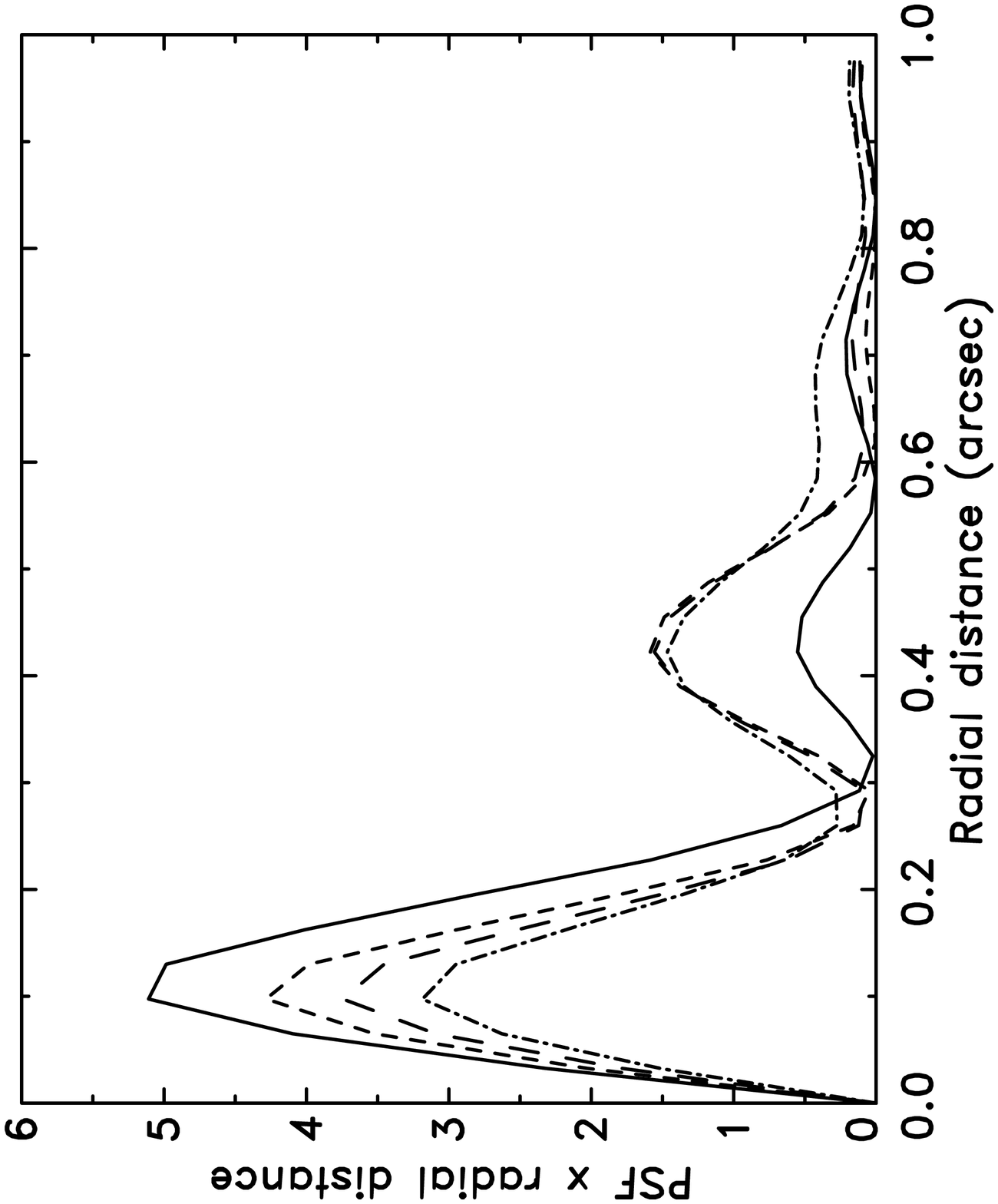}
 \put(87,63){\textcolor{black}{\textbf{b)}}}
\end{overpic}
\begin{overpic}[bb=55 70 570 710,width=0.3\textwidth,angle=-90]{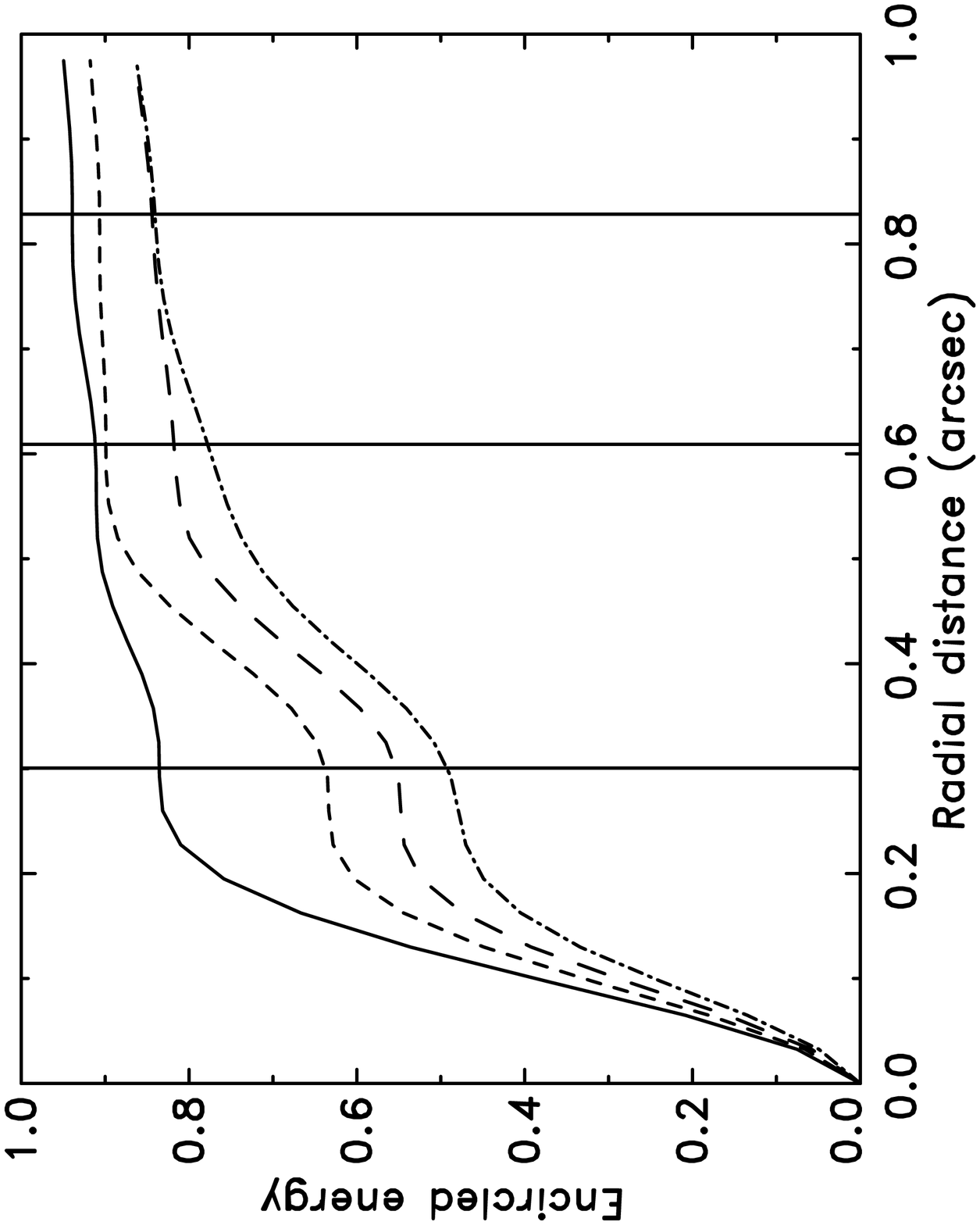}
 \put(87,63){\textcolor{black}{\textbf{c)}}}
\end{overpic}
\begin{overpic}[bb=55 70 570 710,width=0.3\textwidth,angle=-90]{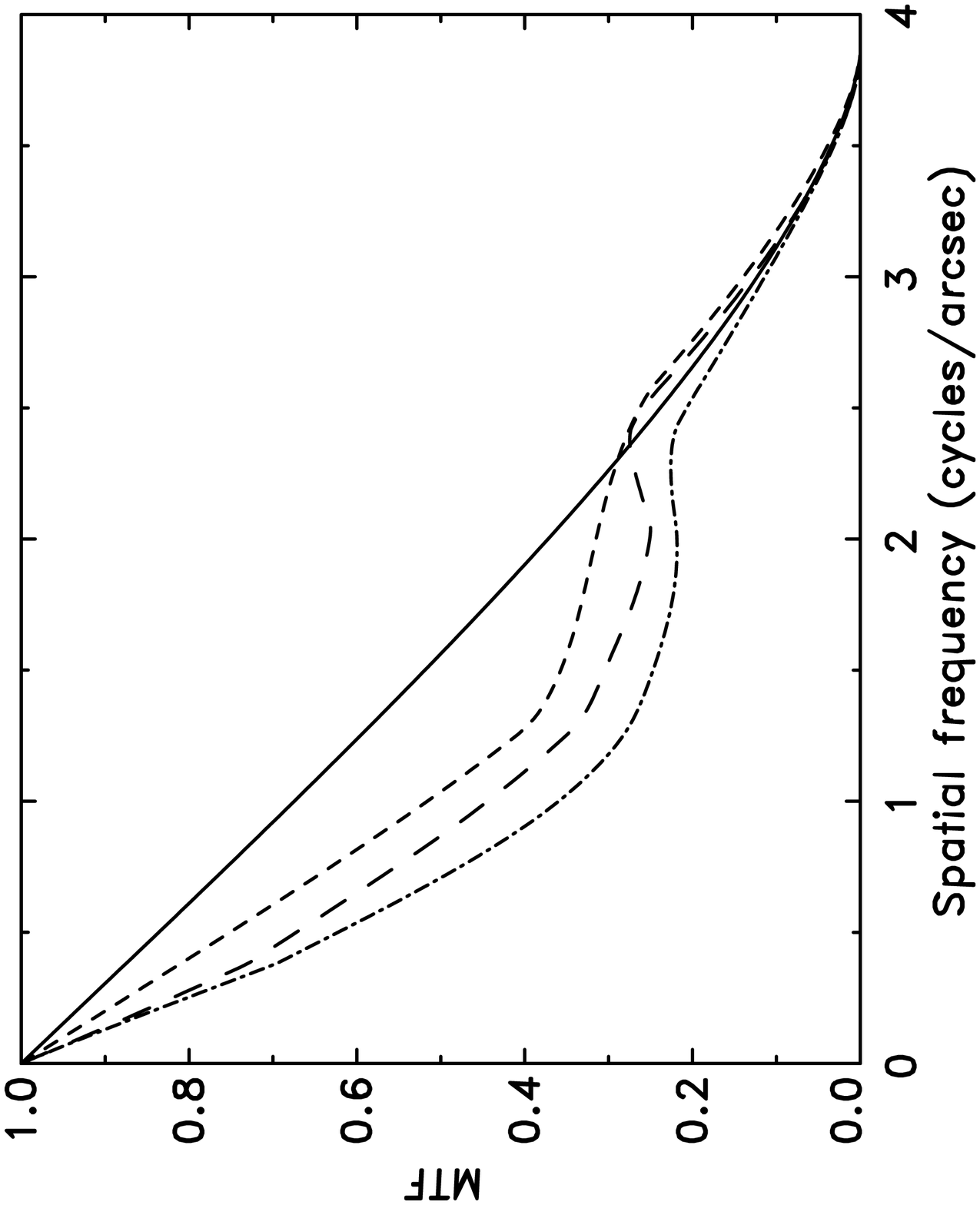}
\put(87,63){\textcolor{black}{\textbf{d)}}}
\end{overpic}
\vspace{0.02cm}\\
\caption{Panel a shows the core of the angle-averaged point spread function (PSF) of Hinode, normalized to unity at its peak, panel 1b the same PSF multiplied by the radial distance $r$ to visualize the contributions from different radial distances. Panel c shows the variation of the encircled energy of the PSF with distance, with the approximate locations of the first three diffraction minima indicated with vertical lines. Panel d shows a cut through the corresponding modulation transfer function. The 4 curves in each panel correspond to Case I (full), Case II (short dashes), Case III (long dashes) and Case IV (dash-dotted), summarized in Table 1.}
\label{fig:Hinode}
\end{figure*}

Figure~1a shows the diffraction core and most of the first diffraction ring of the corresponding angle-averaged PSF's, normalized to unity at the center. As can be seen, the FWHM of the PSF core is not degraded at all by the central obscuration, spider or focus error. On the contrary, the PSF corresponding to Case II-IV are all marginally narrower than the diffraction limited PSF (Case I). This plot therefore firmly establishes that for Case II-IV, the reduced rms granulation contrast calculated by Danilovic et al. is not at all related to a reduction of the spatial resolution of Hinode. Instead, it is related to an increased energy in the first few diffraction rings, at the expense of a corresponding reduction of the energy in the diffraction core. In Fig. 1b, we show the angle-averaged PSF's multiplied with the radial distance $r$, with the integrals of all curves over $r$ properly normalized to unity. The multiplication with $r$ gives a proper impression of the accumulated contributions (i.e, integrated over all azimuthal angles) from a distance $r$ to the total intensity measured at the center pixel. As can be seen, the central obscuration, the spider and the focus error gradually reduce the contributions from within the diffraction core (within the first minimum at about 0\farcs3), and enhances in particular the contributions from the first and second diffraction rings. The only effects of the central obscuration, the spider and the focus error is thus to degrade the image with spatial straylight, whereas the spatial resolution remains intact. Accounting for this straylight in a crude way requires a PSF that has a diameter corresponding to the outer diameter of the first diffraction ring, or about 1\farcs2, but evidently a reasonably accurate representation of the PSF requires extending that outer diameter to that corresponding to at least the second diffraction ring, or about 1\farcs6. 

Figure 1c shows the encircled energy as function of radius for the 4 PSF's. This demonstrates that including only the first diffraction ring corresponds to ignoring more than 20\% of the enclosed energy of the PSF for case IV. Finally, Fig. 1d shows the MTF along a vertical cut to demonstrate consistency with the calculations of \citet{2008A&A...484L..17D}. The agreement is excellent for cases I and II. Note that Danilovic et al. include also the MTF corresponding to the CCD pixel size. This is not done here, explaining the differences for cases III and IV.

Our calculations thus demonstrate and confirm (as should already be clear from the existing literature) that the OTA of Hinode corresponds to a PSF that leads to approximately 35\% spatial straylight and possibly as much as 40\%, if we accept the possibility of 4.7\% additional straylight suggested by \citet{2008A&A...484L..17D}. Since the shape of the core of the PSF is nearly the same for the 4 cases studied, the Strehl values given in Table 1 directly reflect the total amount of energy in the core and therefore also the straylight in the wings. This level of straylight is also obtained by \citet{2009A&A...501L..19M}, determining the Hinode PSF using Mercury transit data. They fit the inferred PSF to 4 Gaussians, where the first Gaussian is stated to represent the diffraction limited PSF and the remaining 3 Gaussians straylight. At long wavelengths, the total contributions of the straylight as given by the weights in their Table 1 can then be concluded to be nearly 38\%. Finally, \citet{2008A&A...487..399W} estimates an approximate upper limit of the Hinode straylight to be 8\% at a distance of 5\arcsec, when comparing to the diffraction limited PSF of Hinode OTA. Adding the straylight provided by the central obscuration and spider as given by $\Delta E_{wing}$ for Case III in Table 1, this corresponds to an upper limit for the total straylight of 37\%, when compared to an unobscured diffraction limited 50-cm telescope.

Most of the Hinode straylight is thus an unavoidable consequence of the pupil geometry with its relatively large 34.4\% central obscuration and the three 40 mm wide spiders. Figure 1c shows that most of the straylight is contained within a diameter of 1\farcs2-1\farcs6, but also that there remains significant contributions well beyond that. The spatial extent of the Hinode straylight thus is comparable to that inferred for the SST. The origin of the straylight is (mostly) different for the two telescopes, since the SST has an unobscured pupil. At present, we cannot make firm statements about the origin of the enhanced straylight in SST data, but the analysis made so far indicates that (small-scale) aberrations must be the dominant source \citep{2012A&A...537A..80L}.

\section{Straylight compensation leads to redshifted profiles for convection}
SF analyse two spectropolarimetric maps obtained with Hinode and point out that compensation for straylight leads to a systematic red-shift of the profiles both in the quiet Sun and in the penumbra. Without straylight compensation of the data, they obtain average quiet Sun blue-shifts of around $-250$~m~s$^{-1}$, but employing 60\% straylight compensation yields an average red-shift of about 100~m~s$^{-1}$. A similar but much smaller effect is found for the penumbra where the average red-shift for the straylight compensated data corresponds to velocities of around 100--200~m~s$^{-1}$. Although not stated explicitly, this suggests to the reader that this is (part of) the explanation for the detection of penumbral convective downflows in the SST data of \citet{2011Sci...333..316S, 2011ApJ...734L..18J}. However, SF neglect to mention that after straylight correction, the SST data used by \citet{2011Sci...333..316S} correspond to an average blue-shift (not redshift!) of $-240$~m~s$^{-1}$ for the quiet Sun (SOM, Table S2) and $-280$~m~s$^{-1}$ for the penumbra (SOM, Table S3). If anything, this should suggest to SF that the measured downflow velocities in the penumbra from the SST data could be {\em underestimated} by over 200~m~s$^{-1}$.

We also remark that the trend of giving increasing red-shifts when compensating for straylight is a direct consequence of a very well-known effect: the convective blue-shift, explained over 30 years ago \citep[e.g.,][]{1978SoPh...58..243B,1982ARA&A..20...61D}. This gives a systematic blue-shift when averaging spectral lines from (convective) bright upflows and dark downflows, as illustrated also by SF. Conversely, compensating for the spatial mixing of the line profiles by straylight must lead to an increased redshift of the averaged measured velocity. This systematic red-shift effect from straylight compensation would be absent without a true correlation between intensity and velocity in the original data in the sense expected for convection. Thus, the systematic redshift effects reported on by SF are precisely what must be expected from small-scale convective processes (SOM, Sect. 1.2). 
 
\section{Discussion and Conclusions}
The paper by SF discusses aspects of straylight compensation of 2D spectroscopy that are to a large extent well-known or trivial to explain. In particular, SF show that overcompensation for straylight can lead to overcompensated velocity fluctuations and that for small-scale convective flows (with bright upflows and dark downflows), such compensation can lead to systematic effects in the velocities in the sense of enhancing downflows.

The problem is that SF use their results to question the downflows detected by \citet{2011Sci...333..316S} and \citet{2011ApJ...734L..18J}, where evidence for convective downflows everywhere in the penumbra are reported for the first time. However, the analysis of SF does not allow such conclusions to be drawn. In particular, SF ignore the constraints imposed from 3D convection simulations on the strength of the straylight PSF and on its spatial extent from the measured minimum umbra intensity. We have shown (SOM, Sect. 1.2) that these constraints are sufficient to establish firmly the existence of such downflows, such that this is a robust result. To question these results requires an analysis with a straylight compensation that uses the same or similar constraints as described in SOM (Sect. 1.2) but yet invalidates the detection of dark downflows in the SST data. This fundamental aspect of the data processing is ignored by SF.

Another fundamental problem is that SF apply straylight compensation to Hinode 2D spectrometric data, coupled with the claim that the straylight of Hinode ``clearly'' is less than 10\%. This misleads the reader into interpreting the enhanced red-shifts obtained by SF as false detections. However, their statement about the Hinode straylight is false. We have shown above that the actual straylight of Hinode must be considered to be at least 35\%, such that many of the red-shifts obtained with the straylight compensation of SF may well be genuine, even though their chosen FWHM of 4\farcs7 for the straylight PSF appears much too large.

It should perhaps be pointed out that the definition of straylight is not without ambiguity in the literature. Straylight is sometimes taken to refer to only the extreme wings of the PSF, or even to purely additive straylight. Nevertheless, in the present context of the straylight from SST and Hinode, no such ambiguity exists. The question raised by SF is whether the straylight inferred for SST data from within 0\farcs6--1\farcs2 of its diffraction peak is real. We have shown in SOM that this must be the case and here that considerable straylight is contributed at similar small scales also from the PSF of Hinode.

We conclude that the analysis presented by SF has no relevance in the context of the straylight compensation applied to SST data by \citet{2011Sci...333..316S} and \citet{2011ApJ...734L..18J}. The claim that the detections of penumbral convective downflows ``could be produced by effects of overcorrection'' therefore lacks scientific justification. This statement of SF is somewhat remarkable also in view of the fact that we have already demonstrated from analysis of \ion{Fe}{I} 630.15~nm data that such convective downflows can be seen (although with reduced strength) even without any straylight compensation \citep[][their Fig. 14]{2012A&A...540A..19S}. 

We finally emphasize that the explanation for why SF did not detect the omni-present downflows seen in the SST data evidently is that these authors did not account for the actual PSF of Hinode, including the straylight of at least 35\%, shown here to exist. By properly accounting for the actual PSF of Hinode, \citet{2013arXiv1307.3668T} find both downflows and the opposite polarity field in exactly the same data as analyzed by \citet{2009A&A...508.1453F} and \citet{2013A&A...550A..97F}, while the latter authors found such downflows mostly in the outer penumbra and much fewer such patches in the mid and (especially) the inner penumbra. Opposite polarity field associated with downflows was previously found also in SST data \citep{2013A&A...553A..63S}.

\begin{acknowledgements}
The anonymous referee is thanked for careful reading of the manuscript and constructive criticism.
\end{acknowledgements}


\end{document}